\documentclass[aps,pre,superscriptaddress,showpacs,showkeys,nofootinbib]{revtex4}

\usepackage{amssymb,amsmath,amsfonts}
\usepackage{graphicx}
\usepackage{color}
\usepackage{url}

\usepackage[latin1]{inputenc}
\usepackage{tikz}
\usetikzlibrary{shapes,arrows}

\newcommand{\bfnabla}{{\boldsymbol\nabla}}
\newcommand{\bfeta}{{\boldsymbol\eta}}
\newcommand{\bfr}{{\boldsymbol r}}

\begin{document}
\title{An $H$ theorem for Boltzmann's equation\\ for the Yard-Sale Model of asset exchange\\
{\small The Gini coefficient as an $H$ functional}
}
\thanks{\copyright 2014, all rights reserved}
\author{Bruce M. Boghosian}
\affiliation{Department of Mathematics, Tufts University, Medford, Massachusetts 02155, USA}
\author{Merek Johnson}
\affiliation{Department of Mathematics, Tufts University, Medford, Massachusetts 02155, USA}
\author{Jeremy A. Marcq}
\affiliation{Department of Mathematics, Boston University, Boston, Massachusetts 02215, USA}
\date{December 1, 2014}
\begin{abstract}
In recent work~\cite{bib:Boghosian1,bib:Boghosian2}, Boltzmann and Fokker-Planck equations were derived for the ``Yard-Sale Model'' of asset exchange.  For the version of the model without redistribution, it was conjectured, based on numerical evidence, that the time-asymptotic state of the model was oligarchy -- complete concentration of wealth by a single individual.  In this work, we prove that conjecture by demonstrating that the Gini coefficient, a measure of inequality commonly used by economists, is an $H$ function of both the Boltzmann and Fokker-Planck equations for the model.
\end{abstract}
\pacs{89.65.Gh, 05.20.Dd}
\keywords{Boltzmann equation, Asset Exchange Model, Yard-Sale Model, H Theorem, Gini coefficient, Pareto distribution}
\maketitle
\tableofcontents

\section{Introduction}

Over one hundred years ago, the Italian economist Vilfredo Pareto~\cite{bib:Pareto} made one of the first empirical studies of the distribution of wealth by undertaking a careful study of land ownership in Italy, Switzerland and Germany.  In the course of this study, he plotted the fraction of economic agents~\footnote{Throughout this paper, we use the term ``economic agent'' to refer to any entity that can hold or exchange wealth.  This includes, but is not limited to individuals, corporations, funds, central banks, etc.  By ``wealth,'' we refer to currency or anything that can be bought or sold with currency.  This includes, but is not limited to, real estate, stock, commodities, factories, etc.} with land holdings worth more than $w$ as a function of $w$.  His studies led him to believe that this function, which we shall denote by $A(w)$ has a universal form.

By its definition, it is clear that $A$ is monotone non-increasing, and that $A(0)=1$ and $\lim_{w\rightarrow\infty}A(w)=0$.  What Pareto observed is that $A(w)$ is approximately equal to one for all $w$ less than a certain cutoff value denoted by $w_{\mbox{\tiny $\min$}}$, and decays as a power law for $w > w_{\mbox{\tiny$\min$}}$.  That is, Pareto's empirical observations led him to conclude that it is approximately true that
\[
A(w) \approx
\left\{
\begin{array}{ll}
1 & \mbox{for $w \leq w_{\mbox{\tiny$\min$}}$}\\
\left(\frac{w_{\mbox{\tiny$\min$}}}{w}\right)^\alpha & \mbox{otherwise.}
\end{array}
\right.
\]
The power $\alpha$ is usually called the Pareto exponent.

To put Pareto's observations in modern terms, we may note that $1-A(w)$ is the cumulative distribution function (CDF) of economic agents, ordered by wealth.  Let us denote the corresponding probability density function (PDF) of agents by $P(w)$, but we shall adopt the convention of normalizing $P$ to the total number of economic agents, rather than to unity, so that $\int_a^b dw\; P(w)$ is the total number of agents with wealth in $[a,b]$, for $0\leq a<b$.

The total number of economic agents is given by the zeroth moment of the PDF of agents,
\[
N = \int_0^\infty dw\; P(w),
\]
and the total amount of wealth in the economy is given by the first moment,
\[
W = \int_0^\infty dw\; P(w) w.
\]
With the above nomenclature established, the function that Pareto plotted is
\[
A(w) = \frac{1}{N}\int_w^\infty dx\; P(x).
\]
From the Fundamental Theorem of Calculus, we see that the PDF of agents is related to the derivative of Pareto's function by~\footnote{Here and throughout, we assume that  derivatives of these quantities with respect to $w$ exist, at least in the distributional sense.}
\[
P(w) = -N\frac{dA(w)}{dw}.
\]
In terms of the PDF of agents, Pareto's empirical observation is then
\begin{equation}
P(w) \approx
\left\{
\begin{array}{ll}
0 & \mbox{for $w \leq w_{\mbox{\tiny$\min$}}$}\\
\frac{\alpha N}{w_{\mbox{\tiny$\min$}}}\left(\frac{w_{\mbox{\tiny$\min$}}}{w}\right)^{\alpha+1} & \mbox{otherwise.}
\end{array}
\right.
\label{eq:ParteoPDF}
\end{equation}

The ubiquity of the Pareto distribution is indeed a very fundamental observation of macroeconomics, but it has resisted a microeconomic explanation for most of the last century.  Recent work~\cite{bib:Boghosian1,bib:Boghosian2} has suggested that the shape of Pareto's curve is due to the fact that wealth distributions satisfy a certain Boltzmann equation.  Before describing this Boltzmann equation, we first examine another common quantification of inequality.

\section{The Lorenz curve}

\begin{figure}
\begin{center}
\includegraphics[bbllx=0,bblly=0,bburx=360,bbury=247,width=.40\textwidth]{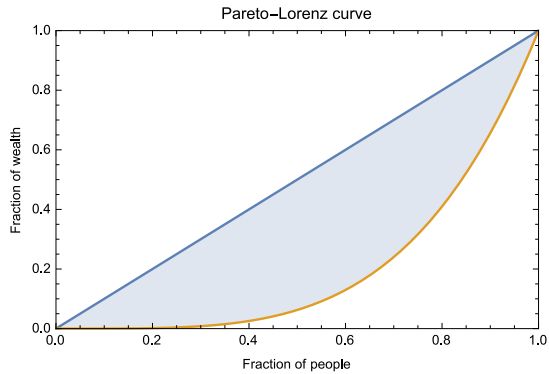}
\end{center}
\caption{{\bf Schematic depiction of a Lorenz curve:}  The cumulative distribution of agents $F(w)$ is plotted on the abscissa, while that of wealth $L(w)$ is plotted on the ordinate.  The parameter $w$ ranges from zero in the lower left corner of the figure, to infinity at the upper right.  The diagonal blue line connecting those two corners corresponds to complete economic equality, in which all agents have the same amount of wealth.  The orange curve is a more realistic depiction of the distribution of wealth in a society.  The fraction of the triangular area under the diagonal that is shaded in blue is the Gini coefficient.}
\label{fig:Pareto-Lorenz}
\end{figure}

In 1906, at approximately the same time as Pareto was making his observations in Europe, the American economist Max O. Lorenz was completing his dissertation work at the University of Wisconsin Madison in which he plotted the cumulative distribution of wealth versus the cumulative distribution of agents.  That is, Lorenz plotted the quantity
\[
L(w) := \frac{1}{W}\int_0^w dx\; P(x) x
\]
versus the quantity
\[
F(w) := \frac{1}{N}\int_0^w dx\; P(x).
\]
Both $L(w)$ and $F(w)$ are monotone nondecreasing, and they increase from zero to one as $w$ goes from zero to infinity.  This graph, which has come to be called the {\it Lorenz curve}, can be plotted in the unit square, as shown schematically in Fig.~\ref{fig:Pareto-Lorenz}.

If all wealth in a society were equally distributed, the Lorenz curve would be the diagonal line, plotted in blue in Fig.~\ref{fig:Pareto-Lorenz}.  This is because any fraction $f\in[0,1]$ of the agents would possess the same fraction $f$ of the total wealth.  In real economies, this curve always lies below the diagonal, more like the orange curve in Fig.~\ref{fig:Pareto-Lorenz}.

To prove that the Lorenz curve can never pass above the diagonal, note first that
\begin{eqnarray}
\frac{dL(w)}{dw} &=& \frac{Pw}{W}\label{eq:dL}\\
\frac{dF(w)}{dw} &=& \frac{P}{N},\label{eq:dF}
\end{eqnarray}
so the slope of the Lorenz curve is given by
\begin{equation}
\frac{dL}{dF} = \frac{N}{W}w,
\label{eq:dLdF}
\end{equation}
which is the ratio of the wealth $w$ corresponding to that point on the curve to the average wealth.  It follows that
\[
\frac{d^2L}{dF^2}=\frac{N}{W}\left(\frac{dF}{dw}\right)^{-1}=\frac{N}{W}\frac{N}{P}\geq 0,
\]
where we have abused notation, in that $L$ is now considered a function of $F$.  We see that $L=F=0$ when $w=0$, and that $L=F=1$ when $w\rightarrow\infty$, and moreover that $L(F)$ is concave up.  It follows that $L(F)$ is bounded above by the diagonal.

Equations (\ref{eq:dL}) and (\ref{eq:dF}) relate the quantities plotted in the Lorenz curve to $P(w)$, and we may also note that
\begin{equation}
A(w) = 1 - F(w).
\label{eq:af}
\end{equation}
Thus, given $N$ and $W$, we see that any of the quantities $A(w)$, $P(w)$, $L(w)$ and $F(w)$ can be derived from any other, so they all contain equivalent information -- which is to say that they all contain essentially complete information about the distribution of wealth in a society.

\section{The Gini coefficient}

In fact, actual data for wealth distributions in the world today is very scant, and economists have to content themselves with much coarser characterizations of wealth inequality than the quantities described above.  One of the most popular of these is due to the Italian statistician and sociologist Corrado Gini, who also worked roughly contemporaneously with Pareto and Lorenz.

The {\it Gini coefficient} $G$ is defined as the ratio of the shaded area in Fig.~\ref{fig:Pareto-Lorenz}, lying between the diagonal segment and the Lorenz curve.  Consequently, $G=0$ when everybody has equal wealth; the limit $G\rightarrow 1$ describes the approach to complete oligarchy.

In terms of the quantities defined above, the Gini coefficient is given by
\[
G = \frac{\int_0^1 dF\;(F-L)}{\int_0^1 dF\;F} = 1 - 2\int_0^1 dF\; L.
\]
Using Eqs.~(\ref{eq:dL}) and (\ref{eq:dF}), we can change integration variables from $F$ to $w$ to find
\begin{equation}
G = 1 - \frac{2}{NW}\int_0^\infty dw\; P(w)\int_0^w dw'\;P(w') w'
= 1 - \frac{2}{N}\int_0^\infty dw\; P(w)L(w).
\label{eq:gini1}
\end{equation}
Alternatively, changing the order of integration yields
\begin{equation}
G = 1 - \frac{2}{NW}\int_0^\infty dw'\;P(w') w'\int_{w'}^\infty dw\; P(w)
= 1 - \frac{2}{W}\int_0^\infty dw\;P(w)A(w) w,
\label{eq:gini2}
\end{equation}
where we have changed the remaining variable of integration from $w'$ to $w$ in the rightmost expression.  Equation (\ref{eq:gini1}) indicates that $G=1-\frac{2}{N}\langle L\rangle$, where the angle brackets denote an average over the PDF $P$.  Likewise, Eq.~(\ref{eq:gini2}) indicates that $G=1-\frac{2}{W}\langle Aw\rangle$.  From these  fundamental relationships between the Gini coefficient and the functions introduced earlier, we see that $G$ is a quadratic functional of $P$, and we may sometimes emphasize this functional dependence by writing it as $G[P]$.

In what follows, we shall need the {\it Fr\'{e}chet derivative} of $G[P]$ with respect to $P$, which is the analog of the gradient in function space.  This is defined by noting that, for any sufficiently well behaved function $\eta(w)$,
\[
\lim_{\epsilon\rightarrow 0}\frac{1}{\epsilon}\left(
G[P+\epsilon\eta] - G[P]
\right)
\]
is a linear functional of $\eta$.  As such, by the Riesz Representation Theorem, it can be expressed as the inner product of $\eta$ and a quantity which we shall denote by $\delta G/\delta P(w)$; that is, for all sufficiently well behaved functions $\eta$, the relation
\[
\left(
\frac{\delta G}{\delta P(w)},
\eta
\right) =
\lim_{\epsilon\rightarrow 0}\frac{1}{\epsilon}\left(
G[P+\epsilon\eta] - G[P]
\right)
\]
defines the Fr\'{e}chet derivative, $\delta G/\delta P(w)$, where the parentheses on the left indicate the standard $L^2$ inner product.  This may be thought of as the infinite-dimensional version of
\[
{\bfnabla g(\bfr)}\cdot{\bfeta} = \lim_{\epsilon\rightarrow 0}
\frac{1}{\epsilon}\left(g(\bfr+\epsilon\bfeta) - g(\bfr)\right),
\]
for a scalar function $g$.

Applying the above definition to Eq.~(\ref{eq:gini2}), it is a straightforward calculation to find
\begin{equation}
\frac{\delta G}{\delta P(w)} =\frac{2}{W}\left[-w+\frac{1}{N}\int_0^w dx\; P(x)(w-x)\right].
\label{eq:frechet}
\end{equation}
Now, if the PDF, $P$, changes in time, it will cause a change in the Gini coefficient given by
\begin{equation}
\frac{dG}{dt}= \int_0^\infty dw \; \frac{\delta G}{\delta P(w)}\frac{\partial P(w)}{\partial t},
\label{eq:dGdt}
\end{equation}
which is an infinite-dimensional version of the ordinary chain rule,
\[
\frac{dg}{dt} = \bfnabla g(\bfr)\cdot\frac{d\bfr}{dt}.
\]
In the next two sections, we shall consider two different dynamical equations for $P$, and use them along with Eq.~(\ref{eq:dGdt}) to place bounds on $dG/dt$.

\section{Background to the Yard-Sale Model of asset exchange}

As noted in the Introduction, though Pareto's law is over a century old, an explanation for it in terms of microeconomic exchange relations between individual agents is still elusive.  The general idea that simple rules for asset exchange might be used to explain wealth distributions appears to be due to Angle~\cite{bib:Angle} in 1986.  Such models have come to be called {\it Asset-Exchange Models} (AEMs) and they typically involve binary transactions between agents involving some increment of wealth $\Delta w$, with rules for which agent gains it and which agent loses it.  The first work applying the mathematical methods of kinetic theory to such models is the paper of Ispolatov, Krapivsky and Redner~\cite{bib:IspolatovKrapivskyRedner} in 1998.  They considered an AEM model in which the agent who loses the wealth is selected with even odds, and in which $\Delta w$ was proportional to the wealth of the losing agent.  Writing in a popular article in 2002, Hayes~\cite{bib:Hayes} noted that in an economy governed by this model, no agent would willingly trade with a poorer agent, and therefore would try to use deception to trade only with wealthier agents.  For this reason, he named the model of Ispolatov et al. the {\it Theft-and-Fraud Model} (TFM).

In the same article mentioned above, Hayes~\cite{bib:Hayes} proposed a variant of the TFM, in which the losing agent is still selected with even odds, but in which $\Delta w$ is proportional to the wealth of the poorer agent, rather than that of the losing agent.  He referred to this as the {\it Yard-Sale Model} (YSM) and noted that it describes an economy in which rich and poor will have no strategic reason not to trade with each other.  The first kinetic theoretical description of the YSM was given by Boghosian in 2014~\cite{bib:Boghosian1,bib:Boghosian2}.  In analogy with \cite{bib:IspolatovKrapivskyRedner}, he derived a Boltzmann equation for the YSM, but then went on to show that in the limit of small $\Delta w$ and frequent transactions, this reduced to a certain nonlinear, nonlocal Fokker-Planck equation.  He presented numerical evidence indicating that the YSM by itself exhibits ``wealth condensation,'' in which all the wealth ends up in the hands of a single oligarch.  He also showed that, when supplemented with a simple model for redistribution, the YSM yields Pareto-like wealth distributions, including a cutoff at low wealth and an approximate power law at large wealth, very reminiscent of Eq.~(\ref{eq:ParteoPDF}).

The YSM can be described by a very simple algorithm.  The version of the algorithm we shall use here is completely equivalent to that described by Boghosian~\cite{bib:Boghosian1,bib:Boghosian2}, though we state it in the following slightly different fashion:  Choose two agents from the population at random to engage in a financial transaction.  Call them agent 1 and agent 2, and denote their respective wealth values by $w_1$ and $w_2$.  The amount of wealth that will be transferred from agent 1 to agent 2 in this transaction is then $\Delta w = \beta\,\mbox{min}(w_1,w_2)$, where $\beta\in(-1,+1)$ is sampled from a {\it symmetric} probability density function denoted by $\eta(\beta)$.  Note that, because $\eta(\beta)$ is symmetric, the two agents both have even odds of winning and losing.

It may seem odd that an algorithm which gives even odds of winning to both agents engaging in a transaction would cause wealth to concentrate, but this is indeed the case.  This was demonstrated by extensive numerical simulations in \cite{bib:Boghosian1}, and there it was conjectured that the time-asymptotic state of $P(w)$ is a generalized function, with support at $w=0$, zeroth and first moments given by $N$ and $W$ respectively, and possibly divergent higher moments.  This corresponds to an oligarchical state with $G=1$.  In this paper, we confirm that conjecture by demonstrating that the Gini coefficient is a monotone increasing Lyapunov functional of the Boltzmann equation for the YSM, and that it reaches a maximum value of $G=1$ in the above-described oligarchical state.

\section{The Boltzmann equation of the Yard-Sale Model}

The Boltzmann equation of the Yard-Sale Model of asset exchange may be written
{\small
\begin{equation}
\frac{\partial P(w)}{\partial t}=
\int_{-1}^{+1} d\beta \; \eta(\beta)
\left\{
-\left[P(w)-\frac{1}{1+\beta}P\left(\frac{w}{1+\beta}\right)\right]
+
\frac{1}{N}\int_0^{\frac{w}{1+\beta}}dx \; P(x) \left[P(w-\beta x)-\frac{1}{1+\beta}P\left(\frac{w}{1+\beta}\right)\right]
\right\},
\label{eq:Boltzmann}
\end{equation}
}
where $\eta(\beta)$ is the symmetric distribution described in the last section.  This was derived by Boghosian~\cite{bib:Boghosian1} using arguments similar to the derivation of the molecular Boltzmann equation and also using a master equation approach, and so we shall not re-derive it in this paper.  Instead, we shall demonstrate that the Gini coefficient is a Lyapunov function for this equation.  That is, we shall show that $G$ is monotone non-decreasing as a consequence of the above dynamics for $P$.  The Lyapunov function of the molecular Boltzmann equation is traditionally called Boltzmann's ``$H$ function.''  Adopting that nomenclature, we will show that the Gini coefficient is an $H$ function for the Yard-Sale Model Boltzmann equation.

Our first task will be to substitute Eq.~(\ref{eq:Boltzmann}) into Eq.~(\ref{eq:dGdt}), and demonstrate that $dG/dt$ thereby obtained is greater than or equal to zero.  We partition this task by rewriting the above as
\[
\left(\frac{\partial P(w)}{\partial t}\right) =
\left(\frac{\partial P(w)}{\partial t}\right)_1 +
\left(\frac{\partial P(w)}{\partial t}\right)_2,
\]
where we have defined
\begin{equation}
\left(\frac{\partial P(w)}{\partial t}\right)_1 =
-\int_{-1}^{+1} d\beta \; \eta(\beta)
\left[P(w)-\frac{1}{1+\beta}P\left(\frac{w}{1+\beta}\right)\right],
\label{eq:Boltzmann1}
\end{equation}
and
\begin{equation}
\left(\frac{\partial P(w)}{\partial t}\right)_2 =
\frac{1}{N}\int_{-1}^{+1} d\beta \; \eta(\beta)
\int_0^{\frac{w}{1+\beta}}dx \; P(x) \left[P(w-\beta x)-\frac{1}{1+\beta}P\left(\frac{w}{1+\beta}\right)\right].
\label{eq:Boltzmann2}
\end{equation}
We shall now show that the corresponding rates of increase of the Gini coefficient given by Eq.~(\ref{eq:dGdt}),
\begin{equation}
\left(\frac{dG}{dt}\right)_j =
\int_0^\infty dw \; \frac{\delta G}{\delta P(w)}\left(\frac{\partial P(w)}{\partial t}\right)_j,
\label{eq:dGdtj}
\end{equation}
are separately greater than or equal to zero for $j=1,2$.

We first consider $(dG/dt)_1$.  Combining Eqs.~(\ref{eq:frechet}), (\ref{eq:Boltzmann1}) and (\ref{eq:dGdtj}), we have
\begin{eqnarray*}
\left(\frac{dG}{dt}\right)_1
&=&
-\frac{2}{W}\int_0^\infty dw\;
\left[
-w+\frac{1}{N}\int_0^w dx\; P(x)(w-x)
\right]
\int_{-1}^1d\beta \;\eta(\beta)
\left[
P(w)-\frac{1}{1+\beta}P\left(\frac{w}{1+\beta}\right)
\right]\\
&=&
\frac{2}{W}\int_{-1}^1d\beta \;\eta(\beta)\int_0^\infty dw\; wP(w)
-\frac{2}{W}\int_{-1}^1d\beta \;\eta(\beta)\int_0^\infty dw\; w\frac{1}{1+\beta}P\left(\frac{w}{1+\beta}\right)\\
& &
+\frac{2}{NW}\int_{-1}^1d\beta \;\eta(\beta)
\int_0^\infty dw\;\frac{1}{1+\beta}P\left(\frac{w}{1+\beta}\right)\int_0^w dx\; P(x)(w-x)\\
& &
-\frac{2}{NW}\int_{-1}^1d\beta \;\eta(\beta)
\int_0^\infty dw\;P(w)\int_0^w dx\; P(x)(w-x).
\end{eqnarray*}
Use the change of variables $u=\frac{w}{1+\beta}$ in the second and third terms, and then change the name of the integration variable from $u$ back to $w$ to obtain
\begin{eqnarray*}
\left(\frac{dG}{dt}\right)_1
&=&
\frac{2}{W}\int_{-1}^1d\beta \;\eta(\beta)\int_0^\infty dw\; wP(w)
-\frac{2}{W}\int_{-1}^1d\beta \;\eta(\beta)\int_0^\infty dw\; w(1+\beta)P(w)\\
& &
+\frac{2}{NW}\int_{-1}^1d\beta \;\eta(\beta)
\int_0^\infty dw\;P(w)\int_0^{(1+\beta)w} dx\; P(x)((1+\beta)w-x)\\
& &
-\frac{2}{NW}\int_{-1}^1d\beta \;\eta(\beta)
\int_0^\infty dw\;P(w)\int_0^w dx\; P(x)(w-x)\\
&=&
-2\int_{-1}^1d\beta \;\eta(\beta)\beta\\
& &
+\frac{2}{NW}\int_{-1}^1d\beta \;\eta(\beta)
\int_0^\infty dw\;P(w)\left[\int_0^{(1+\beta)w} dx\; P(x)((1+\beta)w-x)-\int_0^w dx\; P(x)(w-x)\right].
\end{eqnarray*}
The first term above vanishes because the integrand is odd in $\beta$, integrated from $-1$ to $+1$.  The second term is nonnegative because $\int_0^w dx\; P(x)(w-x)$ is a non-decreasing function of $w$, as follows from
\[
\frac{d}{dw}\int_0^w dx\; P(x)(w-x)=\int_0^w dx \;P(x)\geq 0.
\]
Thus we have demonstrated that
\begin{equation}
\left(\frac{dG}{dt}\right)_1 \geq 0.
\label{eq:dGdt1geq0}
\end{equation}

We next consider $(dG/dt)_2$.  Combining Eqs.~(\ref{eq:frechet}), (\ref{eq:Boltzmann2}) and (\ref{eq:dGdtj}), we have
\begin{eqnarray*}
\left(\frac{dG}{dt}\right)_2
&=&
\int_{-1}^1d\beta \;\eta(\beta)\frac{2}{NW}\int_0^\infty dw\;
\left[-w+\frac{1}{N}\int_0^w dx\; P(x)(w-x)\right] \\
& &
\int_0^{\frac{w}{1+\beta}}dy\; P(y)
\left[P(w-\beta y)-\frac{1}{1+\beta}P\left(\frac{w}{1+\beta}\right)\right]\\
&=&
-\frac{2}{NW}\int_{-1}^1d\beta \;\eta(\beta)
\int_0^\infty dw\;w
\int_0^{\frac{w}{1+\beta}}dy\; P(y)P(w-\beta y) \\
& &
+\frac{2}{NW}\int_{-1}^1d\beta \;\eta(\beta)\frac{1}{1+\beta}
\int_0^\infty dw\;w
\int_0^{\frac{w}{1+\beta}}dy\; P(y)P\left(\frac{w}{1+\beta}\right) \\
& &
+\frac{2}{N^2W}\int_{-1}^1d\beta \;\eta(\beta)
\int_0^\infty dw\;
\int_0^w dx\; P(x)(w-x)
\int_0^{\frac{w}{1+\beta}}dy\; P(y)P(w-\beta y) \\
& &
-\frac{2}{N^2W}\int_{-1}^1d\beta \;\eta(\beta)\frac{1}{1+\beta}
\int_0^\infty dw \int_0^w dx\; P(x)(w-x)
\int_0^{\frac{w}{1+\beta}}dy\; P(y)P\left(\frac{w}{1+\beta}\right).
\end{eqnarray*}
In the first and third terms, swap the order of integration so that $y$ is outermost, and then make the substitution $u=w-\beta y$.  In the second and fourth terms, use the change of variable $u=w/(1+\beta)$, and then swap the order of integration so that $y$ is outermost.  The result is
\begin{eqnarray*}
\left(\frac{dG}{dt}\right)_2
&=&
-\frac{2}{NW}\int_{-1}^1d\beta \;\eta(\beta)
\int_0^\infty dy\;P(y)
\int_{y}^\infty du\;P(u)
(u+\beta y)\\
& &
+\frac{2}{NW}\int_{-1}^1d\beta \;\eta(\beta) (1+\beta)
\int_0^\infty dy\;P(y)
\int_y^\infty du\;P(u) u \\
& &
+\frac{2}{N^2W}\int_{-1}^1d\beta \;\eta(\beta)
\int_0^\infty dy\;P(y)
\int_{y}^\infty du\;P(u)
\int_0^{u+\beta y} dx\;P(x)
((u+\beta y)-x)\\
& &
-\frac{2}{N^2W}\int_{-1}^1d\beta \;\eta(\beta)
\int_0^\infty dy\;P(y)
\int_y^\infty du\;P(u)
\int_0^{(1+\beta)u} dx\;P(x)
((1+\beta)u-x).
\end{eqnarray*}
The integral over $\beta$ can be performed for the first two terms, whereupon they cancel.  Now swap the order of integration in the remaining two integrals so that $x$ is outermost, followed by $y$, and then $u$ to obtain
\begin{eqnarray*}
\left(\frac{dG}{dt}\right)_2
&=&
+\frac{2}{N^2W}\int_{-1}^1d\beta \;\eta(\beta)
\int_0^{\infty} dx\;P(x)
\int_0^{\frac{x}{1+\beta}} dy\;P(y)
\int_{x-\beta y}^\infty du\;P(u)
(u+\beta y-x)\\
& &
+\frac{2}{N^2W}\int_{-1}^1d\beta \;\eta(\beta)
\int_0^{\infty} dx\;P(x)
\int_{\frac{x}{1+\beta}}^\infty dy\;P(y)
\int_{y}^\infty du\;P(u)
(u+\beta y-x)\\
& &
-\frac{2}{N^2W}\int_{-1}^1d\beta \;\eta(\beta)
\int_0^{\infty} dx\;P(x)
\int_0^{\frac{x}{1+\beta}} dy\;P(y)
\int_{\frac{x}{1+\beta}}^\infty du\;P(u)
(u+\beta u-x)\\
& &
-\frac{2}{N^2W}\int_{-1}^1d\beta \;\eta(\beta)
\int_0^{\infty} dx\;P(x)
\int_{\frac{x}{1+\beta}}^\infty dy\;P(y)
\int_y^\infty du\;P(u)
(u+\beta u-x)\\
&=&
+\frac{2}{N^{2}W} \int_{-1}^{+1} d\beta \; \eta(\beta)
\int_{0}^{\infty} dx \; P(x)
\int_{0}^{\frac{x}{1+\beta}} dy \; P(y)\\
& &
\;\;\;\;\;\;\;\;\;\;\;\;\;\;\;\;\;\;\;\;\;\;\;\;\;\;\;\;\;\;\;\;\;\;\;
\left[
\int_{x-\beta y}^{\frac{x}{1+\beta}} du \; P(u)(u-x) +
\beta \int_{x-\beta y}^{\infty} du \; P(u)y -
\beta \int_{\frac{x}{1+\beta}}^{\infty} du \; P(u) \; u
\right] \\
& &
+\frac{2}{N^{2}W} \int_{-1}^{+1} d\beta \; \eta (\beta)
\int_{0}^{\infty} dx \; P(x)
\int_{\frac{x}{1+\beta}}^{\infty} dy \; P(y)
\left[
\beta \int_{y}^{\infty} du \; P(u)y -
\beta \int_{y}^{\infty} du \; P(u)u
\right].
\end{eqnarray*}
In the second term above, note that $\int_{\frac{x}{1+\beta}}^{\infty} dy=\int_{0}^{\infty} dy - \int_{0}^{\frac{x}{1+\beta}}$.  The term with $\int_{0}^{\infty} dy$ then vanishes upon integration over $\beta$, leaving us with
\begin{eqnarray*}
\left(\frac{dG}{dt}\right)_2
&=&
+\frac{2}{N^{2}W} \int_{-1}^{+1} d\beta \; \eta(\beta)
\int_{0}^{\infty} dx \; P(x)
\int_{0}^{\frac{x}{1+\beta}} dy \; P(y)\\
& &
\;\;\;\;\;\;\;\;\;\;\;\;\;\;\;\;\;\;\;\;\;\;\;\;\;\;\;\;\;\;\;\;\;\;\;
\left[
\int_{\frac{x}{1+\beta}}^{x-\beta y} du \; P(u)(x-u)
-\beta \int_y^{x-\beta y} du \; P(u)y
+\beta \int_{y}^{\frac{x}{1+\beta}} du \; P(u) \; u
\right].
\end{eqnarray*}
Next note that because $0\leq y\leq x/(1+\beta)$ and $\beta\geq 0$, it follows that
\[
0 \leq y \leq \frac{x}{1+\beta} \leq x-\beta y \leq x,
\]
so the integral over $u$ from $y$ to $x-\beta y$ can be split into one from $y$ to $x/(1+\beta)$ plus another from $x/(1+\beta)$ to $x-\beta y$, resulting in
\begin{eqnarray*}
\left(\frac{dG}{dt}\right)_2
&=&
+\frac{2}{N^{2}W} \int_{-1}^{+1} d\beta \; \eta(\beta)
\int_{0}^{\infty} dx \; P(x)
\int_{0}^{\frac{x}{1+\beta}} dy \; P(y)\\
& &
\;\;\;\;\;\;\;\;\;\;
\left[
\int_{\frac{x}{1+\beta}}^{x-\beta y} du \; P(u)(x-u)
-\beta \int_y^{\frac{x}{1+\beta}} du \; P(u)y
-\beta \int_{\frac{x}{1+\beta}}^{x-\beta y} du \; P(u)y
+\beta \int_{y}^{\frac{x}{1+\beta}} du \; P(u) \; u
\right]\\
&=&
+\frac{2}{N^{2}W} \int_{-1}^{+1} d\beta \; \eta(\beta)
\int_{0}^{\infty} dx \; P(x)
\int_{0}^{\frac{x}{1+\beta}} dy \; P(y)\\
& &
\;\;\;\;\;\;\;\;\;\;
\left[
\int_{\frac{x}{1+\beta}}^{x-\beta y} du \; P(u)((x-\beta y)-u) + \int_{y}^{\frac{x}{1+\beta}} du \; P(u)(u-y)
\right].
\end{eqnarray*}
The integrands are now manifestly positive, and we may conclude
\begin{equation}
\left(\frac{dG}{dt}\right)_2 \geq 0.
\label{eq:dGdt2geq0}
\end{equation}
The above demonstration is admittedly tedious.  There may be a shorter route to this result, but as of this writing we have not been able to find one.

Combining Eqs.~(\ref{eq:dGdt1geq0}) and (\ref{eq:dGdt2geq0}), we have
\[
\frac{dG}{dt}
=
\left(\frac{dG}{dt}\right)_1 + \left(\frac{dG}{dt}\right)_2
\geq 0,
\]
and so the Gini coefficient is proven to be a Lyapunov functional of the Boltzmann equation.

\section{Fokker-Planck equation for the Yard-Sale Model}

In earlier work, Boghosian~\cite{bib:Boghosian1,bib:Boghosian2} demonstrated that in the limit of small $\beta$, the Boltzmann equation, Eq.~(\ref{eq:Boltzmann}), reduces to a Fokker-Planck equation of the form
\begin{equation}
\frac{\partial P(w)}{\partial t}=\frac{\partial^2}{\partial w^2}\left(\gamma \frac{w^2}{2}C(w)P(w)\right),
\label{eq:FP}
\end{equation}
where
\[
C(w):=\frac{1}{N}\int_0^w dx\; P(x)\left(1-\frac{x^2}{w^2}\right),
\]
and $\gamma$ is a constant.  From its method of derivation, one would expect that $G$ is also a Lyapunov functional of this equation.  Here we demonstrate this directly.

Combining Eqs.~(\ref{eq:frechet}), (\ref{eq:dGdt}) and (\ref{eq:FP}) yields
\begin{eqnarray*}
\frac{dG}{dt}
&=&
\frac{2}{W}\int_0^\infty dw \;
\left[-w+\frac{1}{N}\int_0^w dx\; P(x)(w-x)\right]
\frac{\partial^2}{\partial w^2}\left(\gamma \frac{w^2}{2}C(w)P(w)\right)\\
&=&
-\frac{2}{W}\int_0^\infty dw \;
\left[-1+\frac{1}{N}\int_0^w dx\; P(x)\right]
\frac{\partial}{\partial w}\left(\gamma \frac{w^2}{2}C(w)P(w)\right) \\
&=&
\frac{2}{W}\int_0^\infty dw \; \frac{1}{N}P(w)\gamma \frac{w^2}{2}C(w)P(w)\\
&=&
\frac{\gamma}{NW}\int_0^\infty dw \; [P(w)]^2 C(w)\geq 0,
\end{eqnarray*}
where we have integrated by parts.  The Gini coefficient is thus proven to be a Lyapunov functional of the Fokker-Planck equation.

\section{The time-asymptotic state of the Yard-Sale Model}

It is well known that the equilibrium solution of the molecular Boltzmann equation, namely the Maxwell-Boltzmann distribution, may be found by setting the variation of Boltzmann's $H$ function to zero, under the constraints of fixed mass, momentum and energy.  We may attempt an analogous computation here, but, as conjectured by Boghosian~\cite{bib:Boghosian1}, the equilibrium solution of the Boltzmann equation for the Yard-Sale Model without redistribution is a singular generalized function, $\zeta(w)$.  It is zero for $w\neq 0$, has zeroth moment $N$ and first moment $W$, and its higher moments may not exist.  It is only when redistribution is included that steady-state solutions similar to the Pareto distribution, Eq.~(\ref{eq:ParteoPDF}), are obtained~\cite{bib:Boghosian1}.  Still, it is instructive to see if the variational approach will yield this singular generalized function, so we turn our attention to that problem in this last section.

Using Eq.~(\ref{eq:gini2}) for the Gini coefficient, $G[P]$, we introduce the Lagrange multipliers $\lambda$ and $\mu$ to enforce the constraints
\begin{eqnarray*}
N &=& \int_0^\infty dw\; P(w)\\
W &=& \int_0^\infty dw\; P(w) w,
\end{eqnarray*}
respectively.  We obtain
\begin{eqnarray}
0
&=&
\frac{\delta}{\delta P(w)}\left(G[P] - \lambda\int_0^\infty dw\; P(w) - \mu\int_0^\infty dw\; P(w) w\right)
\nonumber\\
&=&
\frac{2}{W}\left[-w+\frac{1}{N}\int_0^w dx\; P(x)(w-x)\right] - \lambda - \mu w,
\label{eq:var}
\end{eqnarray}
where we have used Eq.~(\ref{eq:frechet}).  Interpreting this in a weak sense, we take the zeroth moment of this with respect to $P$.  Using Eqs.~(\ref{eq:af}), (\ref{eq:gini1}) and (\ref{eq:gini2}), we obtain
\begin{equation}
2(1-G) + \lambda N + \mu W = 0.
\label{eq:mom0}
\end{equation}

Next, differentiating Eq.~(\ref{eq:var}) yields
\begin{eqnarray}
0
&=&
\frac{2}{W}\left[-1+\frac{1}{N}\int_0^w dx\; P(x)\right] - \mu
\nonumber\\
&=&
\frac{2}{W}\left[-1+F(w)\right] - \mu
\nonumber\\
&=&
-\frac{2}{W}A(w) - \mu,
\label{eq:var1}
\end{eqnarray}
where we have used Eq.~(\ref{eq:af}).  Taking the first moment of this with respect to $P$, and using Eq.~(\ref{eq:gini2}), we obtain
\begin{equation}
(1-G) + \mu W = 0.
\label{eq:mom1}
\end{equation}
Also, taking the limit as $w\rightarrow\infty$ of Eq.~(\ref{eq:var1}) yields
\begin{equation}
\mu = 0.
\label{eq:momlim}
\end{equation}

Eqs.~(\ref{eq:mom0}), (\ref{eq:mom1}) and (\ref{eq:momlim}) can be solved simultaneously to yield $\lambda=\mu=0$ and, most importantly, $G=1$.  So, under the dynamics of the Yard-Sale Model, $G$ increases in time and asymptotically approaches the value one, corresponding to a state of ``perfect oligarchy.''

Finally, note that differentiating Eq.~(\ref{eq:var1}) one more time immediately yields $P(w)=0$, valid everywhere expect $w=0$, as expected.

\section{Conclusions}

We have proven that the Gini coefficient $G$ is a Lyapunov function of the Boltzmann equation for the Yard-Sale Model of asset exchange, as well as the Fokker-Planck equation obtained in the limit of small transaction sizes.  We have also shown that the equilibrium distribution, obtained in the time-asymptotic limit, is zero for all $w\neq 0$, and corresponds to $G=1$.

As noted earlier, it is only when the Yard-Sale Model is supplemented with a mechanism for redistribution that steady states similar to the Pareto distribution are found.  With redistribution, however, the Gini coefficient is clearly no longer a Lyapunov functional, since it would be possible to begin with a higher concentration of wealth than that obtained in the time-asymptotic limit.  It may be possible to find a different Lyapunov functional for the Boltzmann or Fokker-Planck equations with a redistribution term, but we leave this as a topic for future study.

\end{document}